\documentclass[journal,12pt,onecolumn,draftclsnofoot,]{IEEEtran}

\usepackage{cite}

\ifCLASSINFOpdf
   \usepackage[pdftex]{graphicx}
\else

\fi

\usepackage{amsmath}
\usepackage[cmintegrals]{newtxmath}
\usepackage{url}
\begin{document}

\title{Quantitative Benchmarks and New Directions for Noise Power Estimation Methods in ISM Radio Environment}

\author{Jakub~Nikonowicz,
        Aamir~Mahmood, Emiliano~Sisinni,~\IEEEmembership{Member,~IEEE}
        and~Mikael~Gidlund,~\IEEEmembership{Member,~IEEE}% <-this % stops a space
\thanks{J. Nikonowicz is with the Faculty of Electronics and Telecommunications, Pozna{\'n} University of Technology, 61-131 Pozna{\'n}, Poland, e-mail: j.nikonowicz@vp.pl.}% <-this % stops a space
\thanks{A. Mahmood and M. Gidlund are with the Department of Information Systems and Technology, Mid Sweden University, 851 70 Sundsvall, Sweden.}% <-this % stops a space
\thanks{E. Sisinni is with the Department of Information Engineering, University of Brescia, 25123 Brescia, Italy}% <-this % stops a space
}

%{Shell \MakeLowercase{\textit{et al.}}: Bare Demo of IEEEtran.cls for IEEE Communications Society Journals}

\maketitle

\begin{abstract} 
Noise power estimation is a key issue in modern wireless communication 
systems. It allows resource allocation by detecting white spectral spaces 
effectively, and gives control over the communication process by adjusting 
transmission power. Thus far, the proposed estimation methods in the 
literature are based on spectral averaging, eigenvalues of sample covariance 
matrix, information theory, and statistical signal analysis. Each method is 
characterized by certain stability, accuracy and complexity. However, the 
existing literature does not provide an appropriate comparison. In this 
paper, we evaluate the performance of the existing estimation techniques 
intensively in terms of stability and accuracy, followed by detailed 
complexity analysis. The basis for comparison is signal-to-noise ratio (SNR) 
estimation in simulated industrial, scientific and medical (ISM) band 
transmission. The source of used background distortions is complex noise 
measurement, recorded by USRP-2932 in an industrial production area. Based on 
the examined solutions, we also analyze the influence of noise samples 
separation techniques on the estimation process. As a response to the defects 
in the used methods, we propose a novel noise samples separation algorithm 
based on the adaptation of rank order filtering (ROF). In addition to simple 
implementation, the proposed method has a very good 0.5 dB root-mean-squared 
error (RMSE) and smaller than 0.1 dB resolution, thus achieving a performance 
that is comparable with the methods exploiting information theory concepts. 
\end{abstract}

\begin{IEEEkeywords}
Noise power etimation, SNR estimation, blind noise separation, rank order filtering.
\end{IEEEkeywords}

% \fi
\IEEEpeerreviewmaketitle

\section{Introduction}
\label{sec:Intro}

\IEEEPARstart{N}{oise} power estimation is an important process associated 
with many different digital signal processing domains. The most common 
include fundamental problems of speech-enhancement and image denoising 
\cite{NPE_Gerkmann, Shrawankar2010, Wu2015, Sijbers2007}. The dependence on the 
knowledge of noise variance also occurs in signal processing for 
segmentation, clustering, noise reduction, statistical inference etc. 
\cite{Sijbers2007}\cite{Makovoz2006}. Noise power estimation has also acquired  a 
particular attention in wireless communication systems for its role in 
cognitive radio (CR) and link adaptation algorithms, which are the main 
motivation of this article.

%With the development of the cognitive radio, the importance of detecting 
%white spaces in a dynamic radio environment becomes an issue of a 
%prime importance. The most commonly used method for evaluating the occupancy 
%of a channel is still a simple energy detection, based on the estimation of 
%noise power. Therefore, utilization of the spectrum is strongly associated 
%with the ability to accurately  estimate power in unoccupied channel \cite{Pandit2017}. 
%One can safely assume that the discussed issue is one of the main 
%requirements of establishing a successful wireless transmission.

In CR the detection of white spaces or spectral occupancy in a dynamic radio 
environment is imperative for opportunistic spectrum access. A simplest and 
commonly used method for evaluating spectral occupancy is energy detection 
(ED), which requires efficient noise power estimation in the band of interest 
for its reliable operation. Many studies show that an uncertainty in noise 
power level estimation severely limits the sensing capability of energy 
detector \cite{Mariani2011Effects}\cite{Tandra2008}. As a result, the 
spectrum utilization is strongly associated with the ability to accurately 
estimate power in unoccupied channel \cite{Pandit2017}. 

With its direct impact on SNR estimation, noise power estimation is also an 
important parameter in SNR dependent link adaptation techniques such as 
adaptive coding 
and modulation, and power control. Link adaptation is a key denominator in 
radio systems for LTE, 5G and WLAN. For dynamic link adaptation, 
a real-time noise power estimator for continuous channel quality monitoring 
is required. 
Using power control as an example, the design parameters of a 
desired estimator can be well understood. Power control has a significant 
impact on communication range, achievable throughput and generated 
interference \cite{slattery2008platform}. Rapid changes in power regulation, 
besides affecting energy efficiency, can be a source of serious impulse noise 
in adjacent channels. On the other hand, a fluent power regulation depends on 
the stability of the estimation results. Therefore, a proper estimation 
method must not only be accurate, but also characterized by possibly low 
standard deviation, which also approximates the resolution of the method.

Unfortunately, the existing literature is mainly focused on the design of new 
solutions for noise power estimation, and lacks a common quantitative 
benchmark for evaluating their performance. In this paper, we first develop a 
simulation model to overcome this gap that allows for carrying out a 
comprehensive and realistic comparison of noise power estimation methods. The 
simulation model reliably substitutes a simple ISM band transmission with 
time-frequency resource allocation. 
%In a noisy channel divided into sub-
%bands, in subsequent time intervals, information signals can occupy resources 
%fragmentary, with different bandwidths and powers. 
Each estimation method, 
independently or by auxiliary algorithms, locates unoccupied parts of the 
spectrum and estimates the noise power. Our model allows to track the 
accuracy of the estimations over consecutive time intervals and collects the 
results to assess the long-term stability of each estimation technique. For 
realistic comparison of the studied estimators, we used real noise as a 
source of background distortion in the developed simulation model. To this 
end, we collected RF background noise traces in an industrial environment 
using National Instrument USRP-2932.

%The most important advantage of the used model is real noise as a source of 
%background distortions. The comparison is based on the measurements made 
%with a National Instrument USRP-2932 placed in the industrial production 
%area. Complex traces of the RF background noise are collected in the 26th 
%IEEE 802.15.4 channel.

%In addition, we propose an innovative combination 
%of a simple ML estimation with novel samples separation technique by adapting . The 
%proposed combination based on the rank order filtering (ROF), improves the 
%accuracy of the ML estimation even by 30\% at the expense of only 14\% higher 
%variance. At the same time, the presented solution reduces the computational 
%cost of samples selection by 50\% with respect to the commonly used Fisher 
%discriminant. Therefore, easy to implement technique based on averaging, the proposed solution 
%achieves the performance comparable with the estimation techniques originating from  
%information theory.

Our second contribution is a novel adaptation of rank order filtering (ROF) 
technique for samples separation. The ROF, in its original form 
\cite{Sijbers2007}, is used to search for local minima and replicate them among the filtering 
area i.e., it determines noise floor only. Our proposal adapts the filtering 
process to analyze energy changes and to identify groups of samples likely to 
be signal bands. This in turn leads us to an innovative combination of a 
simple ML estimation with ROF as samples separation technique. The proposed 
combination improves the accuracy of the estimation by 30\% at the expense of 
only 14\% higher variance. At the same time, the presented solution reduces 
the computational cost of samples separation by 50\% with respect to the 
commonly used Fisher discriminant. Therefore, the proposed solution, easy to 
implement technique based on averaging, achieves the performance comparable 
with the estimation techniques originating from information theory.

The rest of the paper is organized as follows. Section~\ref{sec:RelatedWork} 
gives closely related works and motivation of our work. 
Section~\ref{sec:SystemModel} introduces system model. 
Section~\ref{sec:noiPowEst_Methods} describes commonly used noise power 
estimation methods. Section~\ref{sec:algorithm}, gives a novel 
estimation algorithm. Section~\ref{sec:SimModel} describes  
simulation model. Section~\ref{sec:EstimatorsComparison} compares the 
proposed estimation approach with the literature. Section~\ref{sec:CompComplexity} 
presents computational complexity analysis. Finally, conclusions are 
drawn in Section~\ref{sec:Conclusion}.

%%%%%%%%%%%%%%%%%%%%%%%%%%%%%%%%%%%%%%%%%%%%%%%%%%%%%%%%%%%%%%%%%%%%
%%%%%%%%%%%%%%%%%%%%%%%%%%%%%%%%%%%%%%%%%%%%%%%%%%%%%%%%%%%%%%%%%%%%
%%%%%%%%%%%%%%%%%%%%%%%%%%%%%%%%%%%%%%%%%%%%%%%%%%%%%%%%%%%%%%%%%%%%
%%%%%%%%%%%%%%%%%%%%%%%%%%%%%%%%%%%%%%%%%%%%%%%%%%%%%%%%%%%%%%%%%%%%
%%%%%%%%%%%%%%%%%%%%%%%%%%%%%%%%%%%%%%%%%%%%%%%%%%%%%%%%%%%%%%%%%%%%
\section{Related Work and Motivation}
\label{sec:RelatedWork}

Spectrum usage awareness and optimal exploitation of radio resources is 
particularly important for devices operating in ISM radio bands, where the 
spectrum utilization is not uniform. The devices operating in ISM bands use 
different communication standards, thus operate with different bandwidths and 
transmit powers \cite{jackson2009genesis}. Therein, the accuracy of noise power 
estimation is central to the correct identification and efficient allocation of available 
radio channels. 

Moreover, the ISM band is often used by wireless networks with  
nodes having limited-hardware resources. An important consideration to this end is the complexity of the estimation methods. With low-computing power, 
the nodes must remain aware of the radio environment and effectively schedule 
their transmissions. Therefore, the estimation techniques 
must not only be accurate and stable but also be easy to implement and execute.

Modern researchers have proposed a number of approaches for estimating noise 
power, based on various properties. Most often referred to in the literature are 
maximum likelihood (ML) and minimum variance unbiased (MVU) estimators \cite{Mariani2011Effects, martinez2017reducing, liu2015noise, cordeiro2006ieee, Hamid2013Blind}. Both ML and MVU estimators are based on the principle of power spectrum averaging. However, the range of proposed solutions is much wider and includes, for instance, information theoretic methods as 
Akaike Information Criteria \cite{Sequeira2012On}\cite{Haddad2007Cog} and correlation-based methods e.g., 
covariance based estimator \cite{Hamid2014Sample}\cite{nadakuditi2008sample}. More specialized solutions, for example based on 
statistical relations, can also be found as in minimum mean squared error (MMSE) 
estimator \cite{Yucek2006MMSE}. All of these methods, although dedicated for similar 
applications, differ significantly in terms of accuracy, stability and 
complexity.

However, most of the published descriptions are largely focused on the 
parameters of the individual estimation methods and omitted the mutual relations 
between the existing solutions. As a response to the lack of relevant studies, 
we revisit the current estimation techniques and compare 
them in terms of the most important parameters. According to the authors, it 
is appropriate and necessary to conduct a comprehensive and realistic comparison of 
noise power estimation methods.

%%%%%%%%%%%%%%%%%%%%%%%%%%%%%%%%%%%%%%%%%%%%%%%%%%%%%%%%%%%%%%%%%%%%
%%%%%%%%%%%%%%%%%%%%%%%%%%%%%%%%%%%%%%%%%%%%%%%%%%%%%%%%%%%%%%%%%%%%
%%%%%%%%%%%%%%%%%%%%%%%%%%%%%%%%%%%%%%%%%%%%%%%%%%%%%%%%%%%%%%%%%%%%
%%%%%%%%%%%%%%%%%%%%%%%%%%%%%%%%%%%%%%%%%%%%%%%%%%%%%%%%%%%%%%%%%%%%
%%%%%%%%%%%%%%%%%%%%%%%%%%%%%%%%%%%%%%%%%%%%%%%%%%%%%%%%%%%%%%%%%%%%
\section{System Model}
\label{sec:SystemModel}

Energy analysis can be performed in either time domain or frequency domain. 
Performing time-domain energy analysis requires filter-banks to divide 
spectrum into frequency bands. In frequency domain, spectrum is already 
channelized into subbands by an FFT operation. In this work, 
estimation is performed on a resource block consisting of as a certain number of 
subcarriers in a given number of time frames. Therefore, a frame-by-frame 
processing of time-domain signals is considered, where consecutive frames are 
transformed to the spectral domain by applying FFT.

Let $S_m(n)$ and $W_m(n)$ be the complex signal and noise spectral 
coefficients, with $n$ the frequency-bin index and $m$ the time-frame index. 
The signal and the noise are assumed to be additive in the short-time Fourier 
domain. The complex spectral noisy observation is thus given by
\begin{equation}
X_m(n) = S_m(n) + W_m(n).
\label{eq:sysModel_noisyObs}
\end{equation}

When information signal $S_m$ is absent, the received signal $X_m$ has power 
equal to the variance of the corresponding additive white Gaussian noise. 
Considering noise with zero mean and variance $\sigma_w^2$, and by invoking 
the central limit theorem, the distribution of the noise power can be assumed 
to be $N\left(\sigma_w^2,\frac{\sigma_w^4}{MN}\right)$, where $M$ is the 
number of time frames $m\in[1,M]$ and $N$ is the number of frequency bins 
$n\in[1,N]$. Thus when $S_m(n)$ is present, the SNR can be expressed as
\begin{equation}
\mathrm{SNR} = \frac{\sigma_x^2 - \sigma_w^2}{\sigma_w^2},
\label{eq:sysModel_SNR}
\end{equation}
where $\sigma_x^2$ and $\sigma_w^2$ are the received signal and the noise 
powers respectively.

%%%%%%%%%%%%%%%%%%%%%%%%%%%%%%%%%%%%%%%%%%%%%%%%%%%%%%%%%%%%%%%%%%%%
%%%%%%%%%%%%%%%%%%%%%%%%%%%%%%%%%%%%%%%%%%%%%%%%%%%%%%%%%%%%%%%%%%%%
%%%%%%%%%%%%%%%%%%%%%%%%%%%%%%%%%%%%%%%%%%%%%%%%%%%%%%%%%%%%%%%%%%%%
%%%%%%%%%%%%%%%%%%%%%%%%%%%%%%%%%%%%%%%%%%%%%%%%%%%%%%%%%%%%%%%%%%%%
%%%%%%%%%%%%%%%%%%%%%%%%%%%%%%%%%%%%%%%%%%%%%%%%%%%%%%%%%%%%%%%%%%%%
\section{Noise Power Estimation Methods}
\label{sec:noiPowEst_Methods}

\subsection{Maximum Likelihood Estimator (ML)}
\label{subsec:ML}
The simplest and the most commonly used technique to estimate the noise power is 
ML estimator. The particular popularity of this method 
is due to its extremely simple implementation. In case of white 
Gaussian noise, ML is limited only to the averaging of the power spectrum over 
selected time interval \cite{Mariani2011Effects}\cite{martinez2017reducing}. 
The energy contained in the frequency band of interest can be estimated as
\begin{equation}
\sigma_w^2=\frac{1}{N} \sum_{n=1}^{N}{|X_m(n)|^2}, 
\label{eq:ML_NoisePower}
\end{equation}
where $X_m(n)$ is the $n$-th frequency bin and $N$ is the number of FFT spectral 
components calculated for the noise samples separated from $m$-th time interval.

This simple technique accurately tracks the instantaneous value of noise power. 
Although it reacts quickly to the changes in noise power, each noise burst can 
seriously distort the measurement. As a consequence, this technique is highly 
susceptible to momentary inaccuracies.

\subsection{Minimum Variance Unbiased Estimator (MVU)}
\label{subsec:MVU}
The problem associated with ML estimator can be solved by extending the analysis over more time 
intervals. A solution called as minimum variance unbiased (MVU) estimator is based on 
averaging the noise power over the entire time-frequency block \cite{liu2015noise}. MVU can 
be described as
\begin{equation}
\sigma_w^2 = \frac{1}{MN}\sum_{m=1}^M {} \sum_{n=1}^N|X_m(n)|^2, 
\label{eq:MVU_NoisePower}
\end{equation}
where $X_m(n)$ is the $n$-th out of $N$ frequency bins in the $m$-th FFT 
realization. Also, $M$ is the number of consecutive time intervals processed 
into $M$ spectrum vectors.

The introduced extension results in stabilization of the estimation process in 
longer time period, which makes the MVU estimator more resistant to 
minor disturbances. However depending on the number of averaged intervals, 
the delay in the MVU response to noise changes increases. The delay in response time remains 
one of the major drawbacks of the method.

Note that while processing ML and MVU estimates, it is vital to separate noise 
samples precisely from the received signal. As a result, the accuracy of both methods 
is strongly dependent on the technique of samples separation. The process of 
samples selection can be realized in various ways. In \cite{cordeiro2006ieee}, the authors 
proposed a hybrid technique by combining fine and fast sensing periods. The authors 
in \cite{Hamid2013Blind} used the Fisher's discriminant-based approach to divide the received signal into two 
groups. The other studies, e.g., \cite{Sequeira2012On}\cite{Haddad2007Cog}, refer to the Minimum Description Length 
(MDL) or Akaike Information Criterion (AIC) as an effective method of signal 
source separation.

\subsection{Akaike Information Criterion (AIC)}
\label{subsec:AIC}
Bearing in mind the constraints in ML and MVU estimators, a promising but 
complex estimation process is based on the AIC, which includes the samples 
separation without any knowledge of the received signal \cite{Sequeira2012On}. 
Therefore, it can be assumed as an independent estimation method. The 
estimation technique as described in its original form, i.e. in 
\cite{Sequeira2012On}, is based on the eigenvalues of the covariance matrix. 
However, the estimation process has been significantly simplified in 
\cite{Haddad2007Cog}. For white noise, it replaces the covariance matrix with 
the values of the sorted averaged periodogram. The main expression of AIC 
is given as 
\begin{equation}
\mathrm{AIC}(n)=\big(N-n\big) M \log\big(\alpha(n)\big) + n\big(2N-n\big),
\label{eq:AIC_n}
\end{equation}
where $N$ is the FFT size, $M$ is the number of intervals over which periodogram is 
averaged and $n$ is the number of frequency bins corresponding to the signal 
in $n$-th model. The function $\alpha(n)$ is defined as
\begin{equation}
\alpha(n) = \frac{1}{N-n} \left(\sum_{i=n+1}^N{\lambda_i}\right)\Bigg/\left(
\prod_{i=n+1}^{N}\lambda_i\right)^{1/(N-n)}, 
\label{eq:AIC_alpha_n}
\end{equation} 
where $\lambda_i$ is the power of the $i$-th frequency bin in averaged 
periodogram considered as a substitute of the eigenvalue.

The power bins $\lambda_i$ in the sorted periodogram, up to the index $n_{
\min}$ indicating the minimum value of $\mathrm{AIC}(n)$, are assumed to 
represent the signal. The mean of the remaining periodogram values is considered 
to be an estimate of the noise power
\begin{equation}
\sigma_w^2 = \frac{1}{N - n_{\min}} \sum_{i=n_{\min}}^N \lambda_i.
\label{eq:AIC_NoisePower}
\end{equation}

This simplification makes the AIC-based estimation easier to implement and 
more useful as it remains effective and 
fully independent from the external signal separation techniques. However, it should be 
noted that the effectiveness of the method is based on the averaged 
periodogram created for the entire resource block. Thus it is problematic to 
maintain the reliability based on a single spectrum realization.
 
\subsection{Covariance Based Estimator (CBE)}
\label{subsec:CBE}
Another estimation approach, which departs from the averaging of the power 
samples, is based on the eigenvalues of the sample co-variance matrix. The 
estimation algorithm proposed in \cite{Hamid2014Sample} uses two main 
assumptions. First, the ordered eigenvalue sequence can be separated into 
signal and noise groups according to the bandwidth occupancy ratio. It means 
that for the sample covariance matrix $\mathbf{C}$
\begin{equation}
\mathbf{C} = \frac{1}{N} \mathbf{X} \mathbf{X}^T,
\label{eq:CBE_covMarix}
\end{equation}
in total number of $M$ eigenvalues $S$ of them are considered to represent 
the signal, where $\mathbf{X}$ stands for $N\times M$ sized time-frequency 
block. 
This assumption is correct if the spectral occupancy ratio is equal to $M/S$. 
In such case $(M-S)$ eigenvalues can be assigned to the noise group. 
	
Second, by using descending ordered eigenvalues $\lambda_m$ of the matrix $
\mathbf{C}$, a range of $L$ linearly-spaced expected noise power values 
$\sigma_i^2\in [\sigma_{\min}^2,\sigma_{\max}^2]$ can be generated, where $L$ 
is user defined spacing and
\begin{equation}
\sigma_{\min}^2 = \frac{\lambda_M}{\left(1-\sqrt{M/N}\right)^2}
\label{eq:CBE_sigma_min}
\end{equation}
\begin{equation}
\sigma_{\max}^2= \frac{\lambda_{S+1}}{\left(1-\sqrt{M/N}\right)^2}.  
\label{eq:CBE_sigma_max}
\end{equation}

Assuming that the empirical distribution of the noise eigenvalues denoted as 
$e.d.$ follows the Marcenko Pastuer distribution \cite{nadakuditi2008sample}, goodness of fit testing 
can be used to find the best fitting power value
\begin{equation}
D_l = \left\|e.d. - MP\left((M-S)/N, \sigma_l^2 \right)\right\|_2,
\label{eq:CBE_PoweRValue}
\end{equation}
where $l=1...L$ and $MP$ denotes the Marcenko Pastuer distribution. Following 
that, based on the conducted fit testing, the most accurate variance value is 
selected
\begin{equation}
\sigma_w^2 = \underset{\sigma_l^2}{\operatorname{argmin}}\left(D_l\right). 
\label{eq:CBE_NoisePower}
\end{equation}

Although a solid foundation in mathematics exhibits high reliability, 
the algorithm remains computationally expensive. In addition, it requires 
knowledge of the bandwidth occupancy which means an extension by an auxiliary 
algorithm such as MDL, as studied in \cite{Hamid2014Sample}, is needed.

\subsection{Minimum Mean Squared Error Estimator (MMSE)}
\label{subsec:MMSE}
In MMSE estimation \cite{Yucek2006MMSE}, the noise variance is estimated 
using an MMSE filter at each subcarrier. The filter coefficients are 
calculated using statistics of the noise over last $M-1$ 
intervals. The estimator can be defined as
\begin{equation}
\sigma_w^2 = \sum_{n=1}^N w_n |X_M(n)|^2,
\label{eq:MMSE_NoisePower}
\end{equation}
where $X_M (n)$ are the frequency bins of the last FFT realization and $w_n$ 
are the filter coefficients.

To determine $w_n$ coefficients, it is necessary to calculate the variance $
\sigma_n^2$ for each $n$-th subcarrier
\begin{equation}
\sigma_n^2= \frac{1}{M-1}\sum_{m=1}^{M-1} |X_m(n)|^2. 
\label{eq:MMSE_variance_n}
\end{equation}

Following that, the correlation of variance in the frequency dimension needs to 
be calculated. The correlation expressed as
\begin{equation}
r(\Delta)=E\{\sigma_n^2 \sigma_{n+\Delta}^2\}
\label{eq:MMSE_Corr}
\end{equation}
is used directly to generate the covariance matrix $\mathbf{C}$, where
\begin{equation}
C(n,m)=r(n-m).
\label{eq:MMSE_C}
\end{equation}
Based on matrix $\mathbf{C}$, vector $\mathbf{r}$ and a unit matrix $
\mathbf{I}$, the coefficients vector $\mathbf{w}$ can be calculated as
\begin{equation}
\mathbf{w}=\left(\mathbf{C}+\mathbf{r}(0)\mathbf{I}\right)^{-1} \mathbf{r}.
\label{eq:MMSE_Coeff}
\end{equation}

The commonly used approach for noise power estimation in multichannel systems 
is however based on $X_m (n)$ belonging to the distribution $N\left(0, \sigma_
n^2\right)$. This assumption requires subtraction of noisy received symbols by 
the best hypothesis of the noiseless ones. The subtraction operation requires
the pool of transmitted symbols to be known, and therefore MMSE cannot 
be considered fully blind. However, if these requirements are fulfilled, 
MMSE can be a convenient alternative to MVU, AIC and CBE. Unlike the other 
methods, it uses all the received samples to estimate noise power. Also, it 
does not require white noise distribution across the analyzed spectrum.

%%%%%%%%%%%%%%%%%%%%%%%%%%%%%%%%%%%%%%%%%%%%%%%%%%%%%%%%%%%%%%%%%%%%
%%%%%%%%%%%%%%%%%%%%%%%%%%%%%%%%%%%%%%%%%%%%%%%%%%%%%%%%%%%%%%%%%%%%
%%%%%%%%%%%%%%%%%%%%%%%%%%%%%%%%%%%%%%%%%%%%%%%%%%%%%%%%%%%%%%%%%%%%
%%%%%%%%%%%%%%%%%%%%%%%%%%%%%%%%%%%%%%%%%%%%%%%%%%%%%%%%%%%%%%%%%%%%
%%%%%%%%%%%%%%%%%%%%%%%%%%%%%%%%%%%%%%%%%%%%%%%%%%%%%%%%%%%%%%%%%%%%
\section{Proposed Separation Algorithm}
\label{sec:algorithm}
\begin{figure}[!h]
	\centering
		\includegraphics[width=0.6\linewidth]{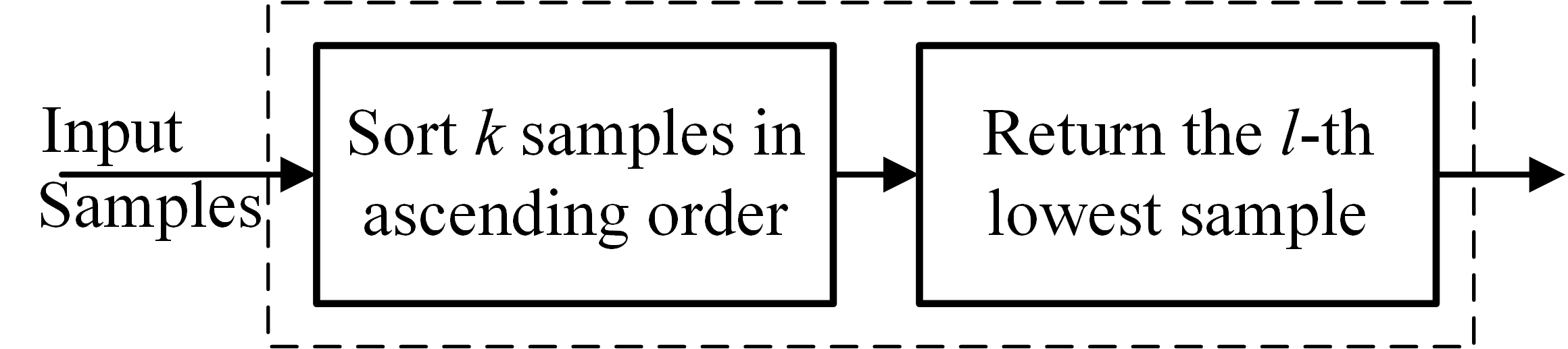}
%\vspace{-18pt}
	\caption{A block diagram of rank order filter $R(k,l)$}
	\label{fig:ROF}
	%\vspace{-14pt}
\end{figure}

As mentioned earlier, in order to successfully perform ML estimation and 
strongly related with it MVU, it is necessary to use signal separation techniques such 
as the Fisher discriminant, MDL or AIC. However, it should be noted that the 
separation of signal is used in many other areas of signal processing. 
Thus the range of proposed solutions is very wide. However to perform ML or MVU in 
simple wireless network nodes effectively, it is 
important for the separation technique to remain simple. In this paper, we 
propose a low-complexity algorithm based on the adaptation of 
Rank Order Filtering (ROF, Fig.~\ref{fig:ROF}).

\begin{figure}[!ht]
	\centering
		\includegraphics[width=0.42\linewidth]{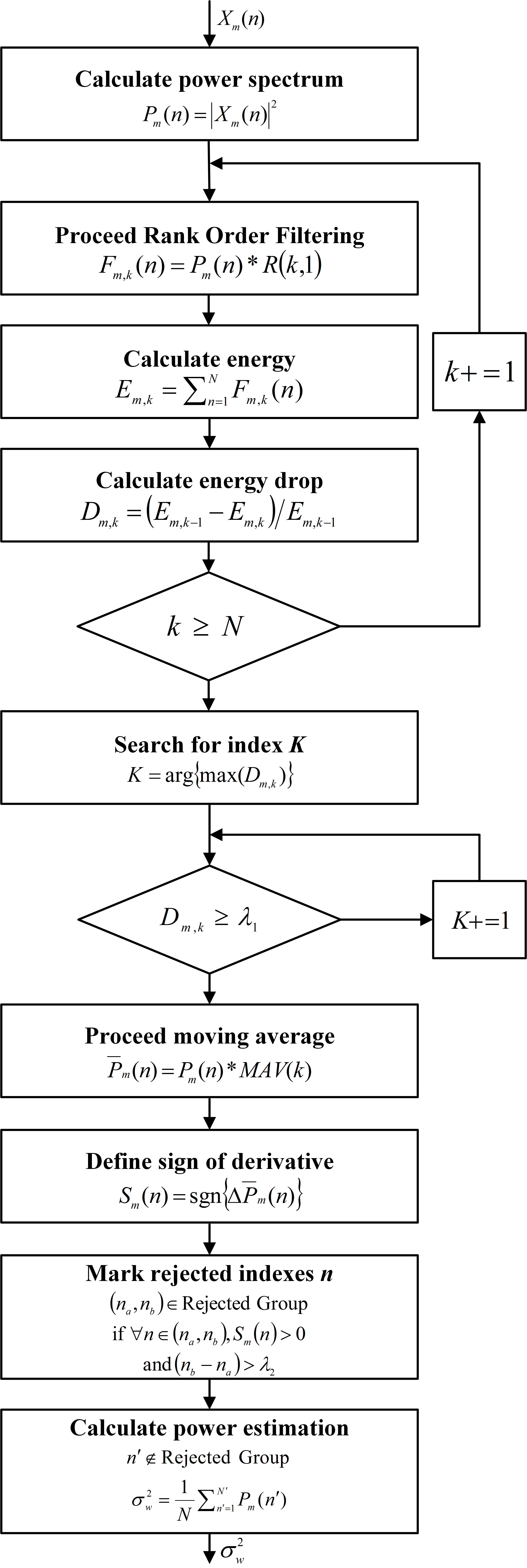}
%\vspace{-18pt}
	\caption{Noise power estimation algorithm using ROF-based sample separation.}
	\label{fig:NPE_ROF}
	%\vspace{-14pt}
\end{figure}

The ROF is originally dedicated to impulse noise reduction and automatic 
noise floor estimation \cite{ready1997automatic}. We modify the ROF method to 
analyze power drop during the filtering process, which we utilize to separate 
spectral samples and perform ML or MVU on the selected noise group. A 
step-by-step operation of the proposed algorithm is shown in 
Fig.~\ref{fig:NPE_ROF}.

The separation process starts with the iterative erosive filtering $R(k,1)$ performed on 
the power spectrum vector $P_m(n)$. The initial value of $k$ is set to 2 and $k$ 
is increased in each iteration until it reaches the size of spectrum vector 
$N$. After filtering the entire input vector with filter size $k$, energy $E_{m,k}$ of the output signal 
$F_{m,k}(n)$ is calculated. The consistent increase in the size of the filter $k$ allows 
to find the size $K$, after which the energy decrease $D_{m,k}$ in relation 
to that of the previous filtering process is highest.

The found $K$ indicates the bandwidth of the signal with the highest power. To 
sensitize the algorithm also for possible signals with wider bands carrying less 
power, index $K$ is iteratively increased until analyzed energy decrease $D_{m,k}$ 
achieves a value lower than the assumed threshold $\lambda_1$. The resulting filter size 
$K$ defines the broadest signal band in the analyzed spectrum. 

Using a $K$-sized moving average on the spectrum $P_m(n)$ results in a 
smoothed spectrum $\bar{P}_m(n)$ in which the long-rising edges of the 
signals are feasible indicators of the currently occupied subbands.
\begin{figure}[!t]
	\centering
		\includegraphics[width=0.7\linewidth]{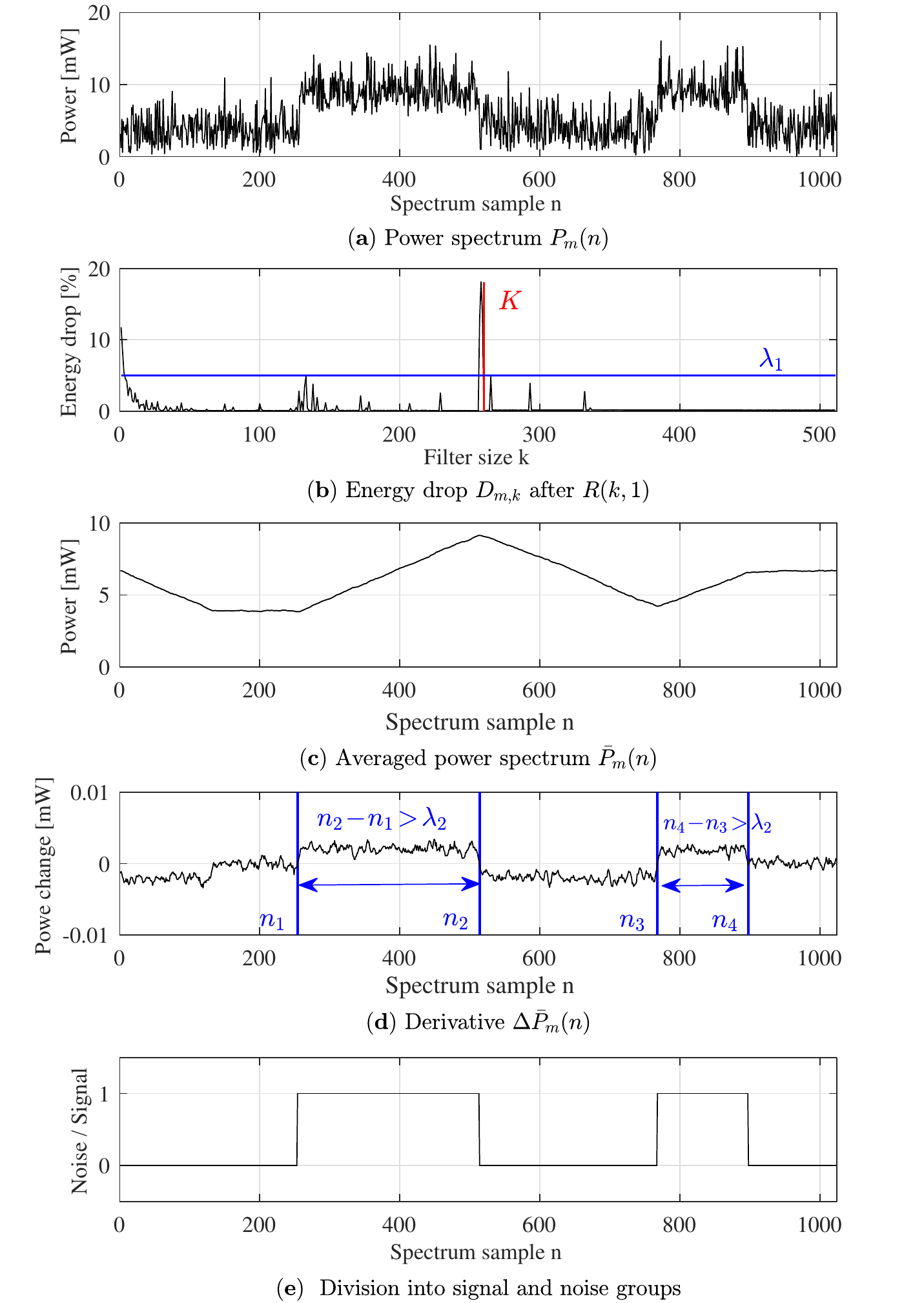}
		%\vspace{-18pt}
	\caption{Process of noise samples separation using modified ROF}
	\label{fig:ROF_Algo}
	%\vspace{-14pt}
\end{figure}
%0.1243220338983052
	
The identification of signal bands is based on the sign of a derivative 
function $\Delta\bar{P}_m(n)$. The areas $(n_a,n_b)$ of the averaged 
spectrum for which derivative has positive values and which are wider than the 
assumed threshold $\lambda_2$ indicate signal subbands. These areas of the 
spectrum can be rejected. The remaining samples $P_m(n)$ indicated by the 
indexes $n^\prime$, not assigned to the signal group, can be used to estimate 
the noise power $\sigma_w^2$.

Fig.~\ref{fig:ROF_Algo} shows the operation of separation algorithm on an 
exemplary signal frame. Fig.~\ref{fig:ROF_Algo}(a) depicts the input power 
vector, which represents a noisy channel with two signal subbands. 
Fig.~\ref{fig:ROF_Algo}(b) presents percentage energy drop after erosive filtering in 
terms of the increase of the filter size $k$. The desired size $K$ is 
marked in red, while assumed threshold $\lambda_1$ (equal to 5\%) is 
marked as a blue line. Fig.~\ref{fig:ROF_Algo}(c) shows the power spectrum trace 
after $K$-size moving average. Results of the derivative calculation from 
the averaged power spectrum are given in Fig.~\ref{fig:ROF_Algo}(d). 
The blue lines indicate positive areas $(n_1,n_2)$ and $(n_3,n_4)$ wider than the
assumed threshold $\lambda_2$ (equal to 5\% of the total bandwidth). Positive 
areas between the lines are marked as a signal group while the rest of the 
spectrum is assumed to represent noise. The separation into signal and 
noise group is shown in Fig.~\ref{fig:ROF_Algo}(e). The samples from 
Fig.~\ref{fig:ROF_Algo}(a) with indexes marked on Fig.~\ref{fig:ROF_Algo}(e) as 
signal should be rejected. Rest of the samples should be used to estimate the 
noise power.

Based on the group of samples classified as noise, effective ML estimation 
for a single time interval can be performed. Therefore, performing the 
separation process also for the following intervals allows for MVU 
estimation.

Note that the thresholds $\lambda_1$ and $\lambda_2$ were chosen empirically. 
A change in these thresholds may effect the efficiency of the algorithm. However, the 
optimization of $\lambda_1$ and $\lambda_2$ remains open for further research.

%%%%%%%%%%%%%%%%%%%%%%%%%%%%%%%%%%%%%%%%%%%%%%%%%%%%%%%%%%%%%%%%%%%%
%%%%%%%%%%%%%%%%%%%%%%%%%%%%%%%%%%%%%%%%%%%%%%%%%%%%%%%%%%%%%%%%%%%%
%%%%%%%%%%%%%%%%%%%%%%%%%%%%%%%%%%%%%%%%%%%%%%%%%%%%%%%%%%%%%%%%%%%%
%%%%%%%%%%%%%%%%%%%%%%%%%%%%%%%%%%%%%%%%%%%%%%%%%%%%%%%%%%%%%%%%%%%%
%%%%%%%%%%%%%%%%%%%%%%%%%%%%%%%%%%%%%%%%%%%%%%%%%%%%%%%%%%%%%%%%%%%%
\section{Simulation Model}
\label{sec:SimModel}
A fair performance comparison of the estimation methods requires a common 
signal platform. The adopted model should reliably approximate transmission 
in ISM band, where various heterogeneous devices occupy resources using 
different bandwidths and powers. In addition, the used model should also 
simulate variability of the tracked noise power and allow analysis of the 
responses to SNR changes in the band of interest.
\begin{figure}[!t]
	\centering
	\vspace{-15pt}
		\includegraphics[width=0.7\linewidth]{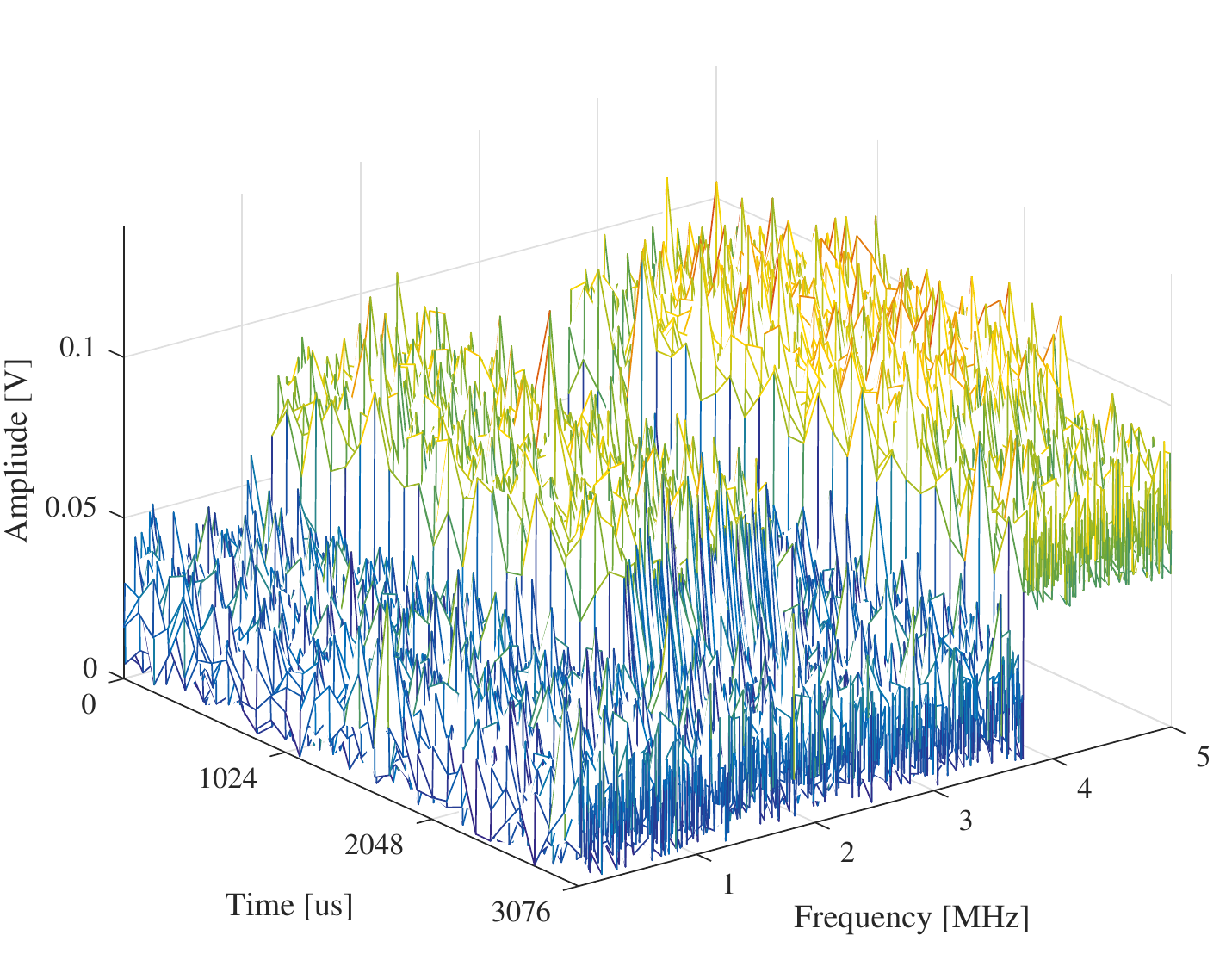}	
%\vspace{-18pt}
	\caption{Time-frequency resource block. Subband spectrum analysis.}
	\label{fig:ResourceBlock}
	%\vspace{-14pt}
\end{figure}

We developed a simulation model that handles the requirements of ISM radio 
environment. In our simulations, the subject of analysis is a time-frequency 
block (Fig.~\ref{fig:ResourceBlock}). It consists of $N$-point FFTs 
calculated for $M$ non-overlapping time intervals. The resource block is 
divided into four equal frequency subbands, to which the signal with 
regulated band fulfillment and power can be added.

\begin{figure}[!h]
	\centering
		\includegraphics[width=0.7\linewidth]{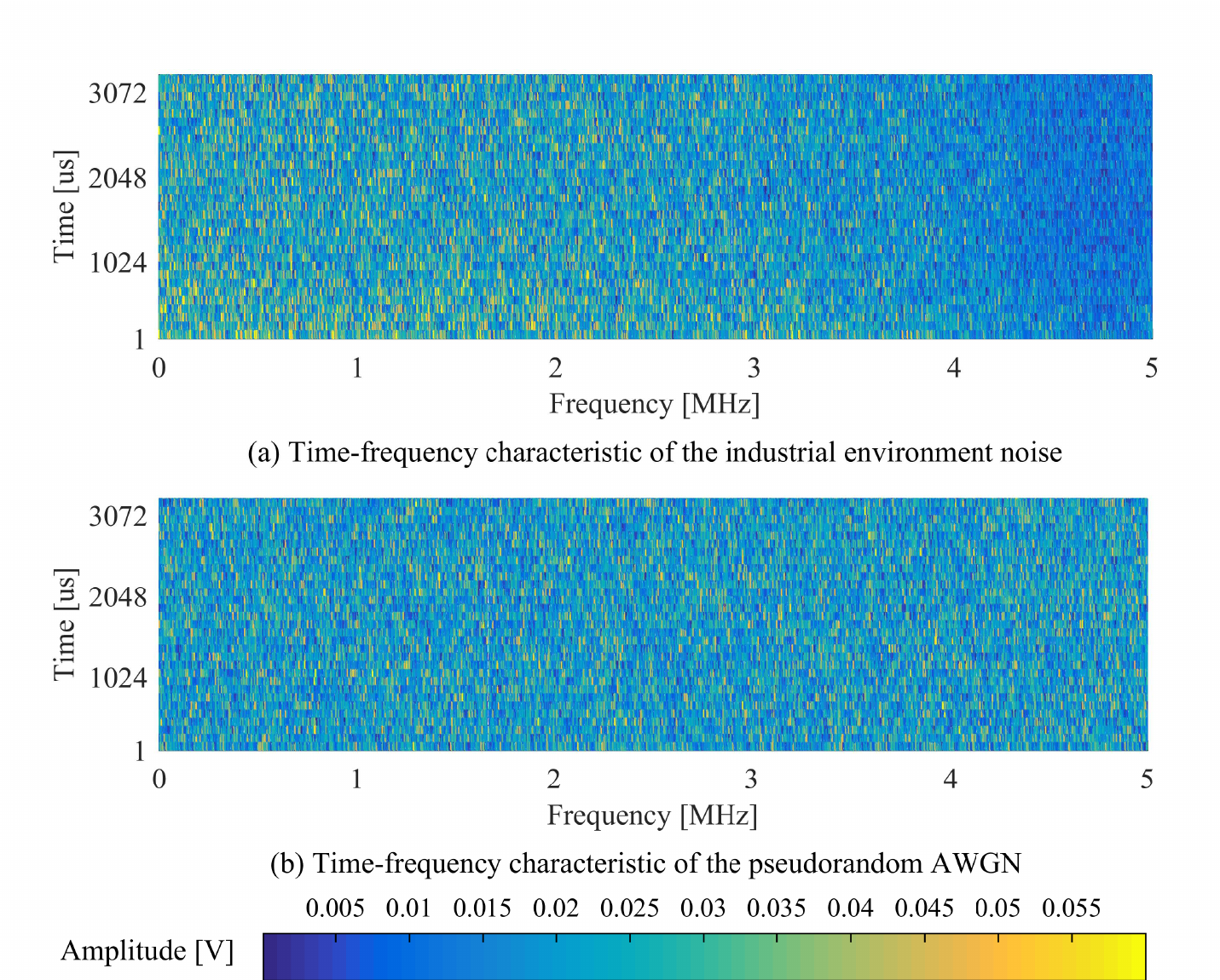}
%\vspace{-18pt}
	\caption{Comparison of a real industrial environment noise and a pseudorandom white Gaussian noise.}
	\label{fig:NoiseComp}
	%\vspace{-14pt}
\end{figure}

The signal collected by a radio receiver may significantly differ from the 
pseudorandom sequences adjusted to the theoretical assumptions of the white 
Gaussian noise. The latter is the most common approach followed in 
performance studies. However, the real noise characteristics include uneven 
distribution of the noise among frequencies and distortions caused by the 
receiver's limitations. Therefore, the most important advantage of the 
presented simulation model is the use of real noise traces as a source of 
distortion (see Fig.~\ref{fig:NoiseComp}).

\begin{figure}[!t]
	\centering
		\includegraphics[width=0.7\linewidth]{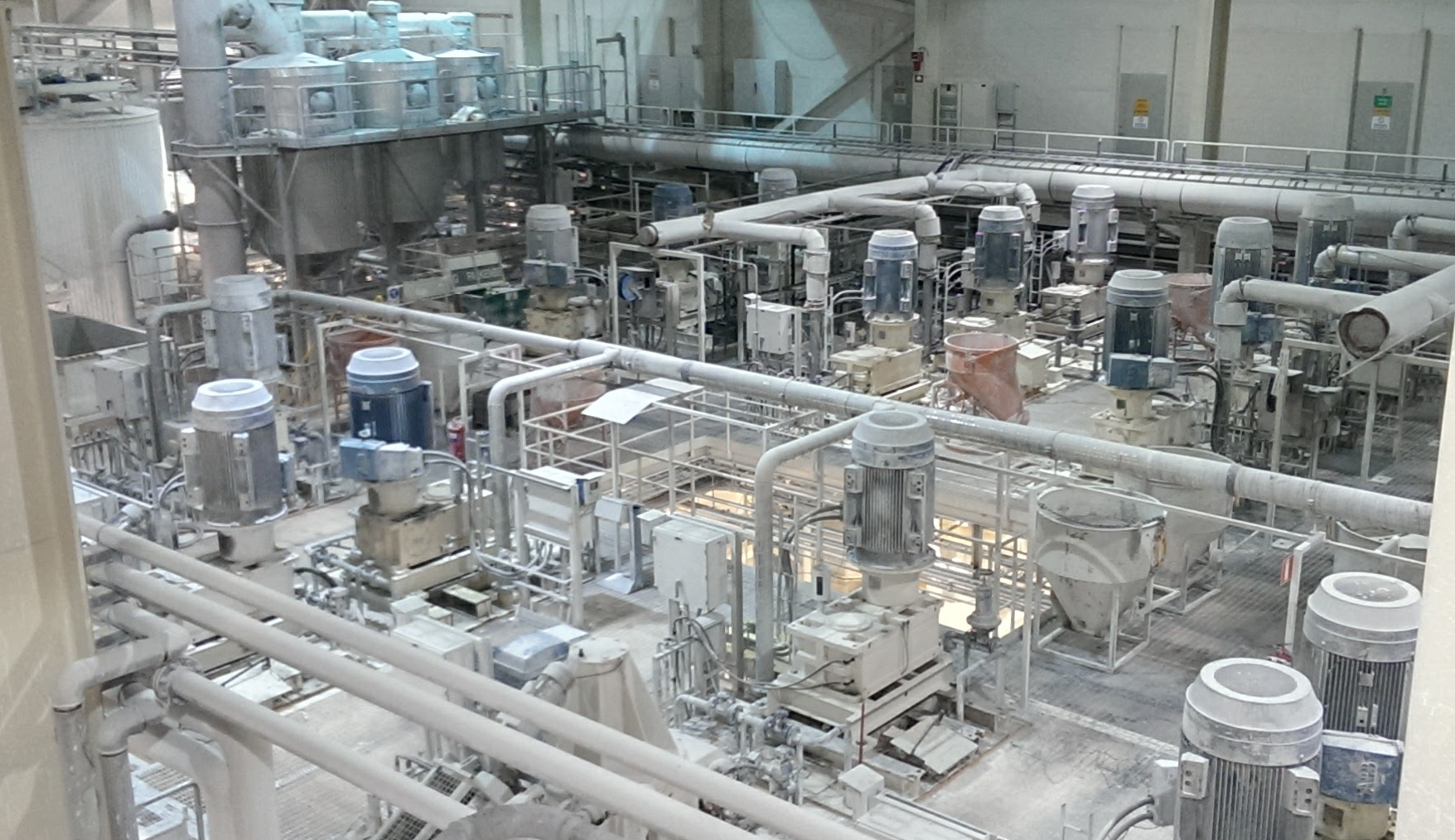}		
%\vspace{-18pt}
	\caption{Industrial environment where the traces are collected.}
	\label{fig:Industry}
	%\vspace{-14pt}
\end{figure}

The real noise traces are the collection of I/Q data of the RF 
background noise in channel 26 of IEEE 802.15.4. We collected the traces in 
an industrial production area using National Instrument USRP-2932. Fig.~\ref{fig:Industry} shows a layout of the industrial plant. The 
channel bandwidth is 5 MHz with the center frequency at 2480 MHz. The 
sampling frequency is set to 10 Msamples/s and the I/Q samples are stored in 
32-bit single precision format.

The simulation model determines the reference noise power based on the 
variance $\sigma_w^2$ of the noise time samples collected in $M$ consecutive time 
intervals. The adjustment of the reference noise power is obtained by rescaling the 
specified variance $\sigma_w^2$. 

The examined estimation scenario also required to append a deterministic 
signal with given bandwidth and power. Due to the simplicity of the 
parameters manipulation, we used a rectangular function in the 
frequency domain, defined as
\begin{equation}
S_m(n) = A_m \mathbf{rect}\left(\frac{n-n_c}{n_a-n_b}\right),
\label{eq:SimModel_rect}
\end{equation}
where $A_m$ is the amplitude of a rectangular function, $n_a$ and $n_b$ are the  
boundary indexes of the spectral bins of the rectangular function and, $n_c=\left(n_a+n_b \right)/2$ is the central bin of the
function $\mathbf{rect}(\cdot)$. Therefore the required spectral energy of a signal can easily be
defined as
\begin{equation}
E_m = A_m^2\left(n_a-n_b\right),
\label{eq:SimModel_specEner}
\end{equation}
which can be regulated by either changing the bandwidth or amplitude.

%%%%%%%%%%%%%%%%%%%%%%%%%%%%%%%%%%%%%%%%%%%%%%%%%%%%%%%%%%%%%%%%%%%%
%%%%%%%%%%%%%%%%%%%%%%%%%%%%%%%%%%%%%%%%%%%%%%%%%%%%%%%%%%%%%%%%%%%%
%%%%%%%%%%%%%%%%%%%%%%%%%%%%%%%%%%%%%%%%%%%%%%%%%%%%%%%%%%%%%%%%%%%%
%%%%%%%%%%%%%%%%%%%%%%%%%%%%%%%%%%%%%%%%%%%%%%%%%%%%%%%%%%%%%%%%%%%%
%%%%%%%%%%%%%%%%%%%%%%%%%%%%%%%%%%%%%%%%%%%%%%%%%%%%%%%%%%%%%%%%%%%%
\section{Comparison of the Estimation Methods}
\label{sec:EstimatorsComparison}

Comparison is carried out for the real noise realization rescaled to 1mW 
average power. The simulated scenario assumes spectrum division into four 
subbands and the presence of a deterministic signal in a single subband occupied 
in 100\% of width. Thus, $n_a$ and $n_b$ are set to 2.5 MHz and 3.75 MHz 
respectively. The regulated SNR is set to 0 dB and -3 dB modified by signal 
amplitude equal to 63.2 mV and 44.7 mV in particular cases. 

\begin{figure}[!t]
	\centering
	\vspace{-5pt}
		\includegraphics[width=0.65\linewidth]{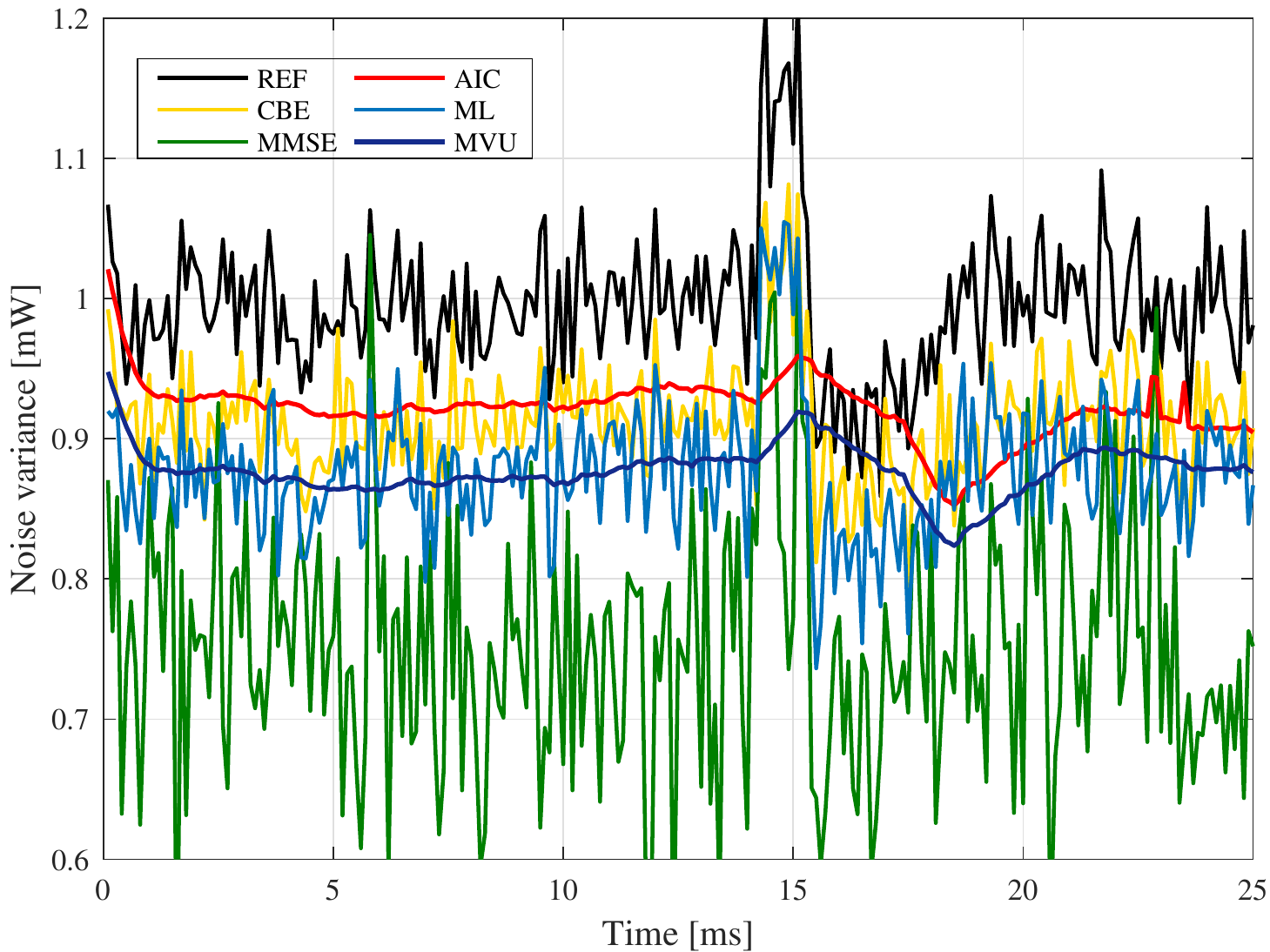}
		\vspace{-12pt}
	\caption{Comparison of the noise power estimation for different estimation methods.}
	\label{fig:NPE_EM}
	%\vspace{-14pt}
\end{figure}

\begin{figure}[!ht]
	\centering
	\vspace{-5pt}
		\includegraphics[width=0.65\linewidth]{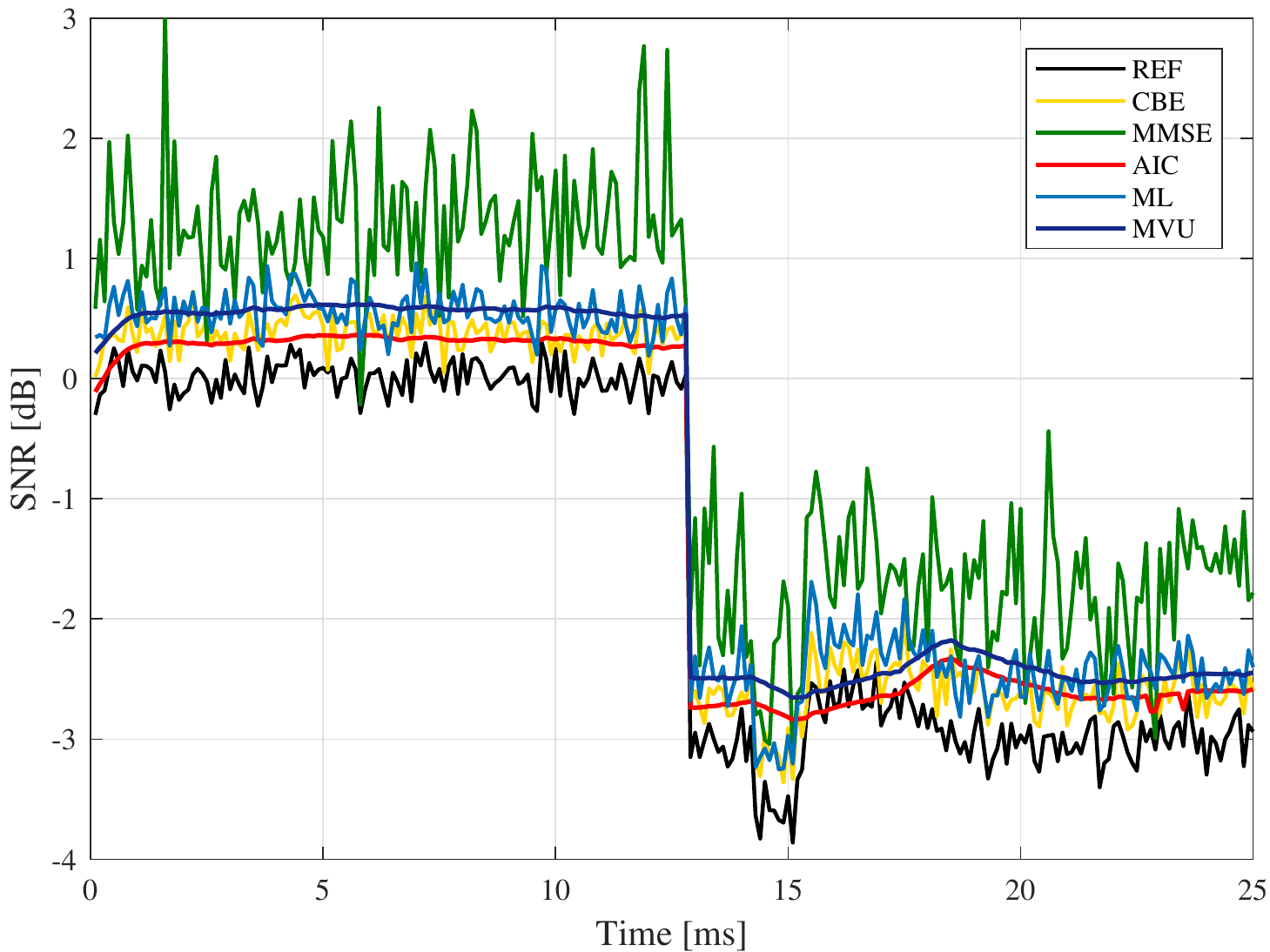}
		\vspace{-12pt}
	\caption{Comparison of SNR estimation for different estimation methods.}
	\label{fig:SNR_EM}
	%\vspace{-14pt}
\end{figure}

In our study, we analyze the ability of estimation methods to track noise 
power in a variable environment (Fig.~\ref{fig:NPE_EM}). However, 
the main focus is on the utility of each method in SNR mapping 
(Fig.~\ref{fig:SNR_EM}). Each technique is evaluated in terms of accuracy by the root-
mean-squared error analysis and stability expressed by standard deviation (see Table~\ref{tab:Tab1}).

For comparison, the MMSE method is adapted for the purpose of the 
blind analysis. The knowledge about the shapes of the transmitted symbols is 
replaced by the reduction in the average value of the frequency bins over time in 
appropriate subbands. Thus, the processed signal is reduced to the 
theoretical distribution $N\left(0,\sigma^2\right)$, as postulated 
in \cite{nadakuditi2008sample}.
	
In case of real noise, which does not completely fulfill the theoretical 
assumptions of the normal distribution, MMSE results in the underestimation 
of the noise power. This leads to overestimation of the SNR and as a 
result may cause too small power increase in the channel to keep 
transmission effective. 
	
A similar overestimation tendency is exhibited by CBE method. However, the detection based 
on the eigenvalues is characterized by significantly higher stability. In 
comparison with other, simpler techniques, CBE does not provide significant 
improvement in estimation accuracy.
	
Among the analyzed estimation techniques ML, MVU and AIC can be considered a 
group of related solutions, i.e. estimators based on the averaging of selected power 
samples. In this group, AIC undoubtedly deserves particular 
attention. AIC, despite replacing the eigenvalues with averaged 
periodogram, reliably approximates the noise power. Thus allows steady and 
accurate SNR estimation, and remains unsusceptible to sudden power 
fluctuations.
\bgroup
\def\arraystretch{1.3}%  1 is the default, change whatever you need
\begin{table}[!h]
%\small
\centering
\vspace{-10pt}
 \caption{SNR Estimation Accuracy and Stability for Different Estimation Methods}
\vspace{-10pt}
\label{tab:Tab1}
\centering
\begin{tabular}{l|ccccc}
\noalign{\hrule height 1pt}
 \textbf{Parameter} & \textbf{CBE} & \textbf{MMSE} & \textbf{AIC} & \textbf{ML} & \textbf{MVU}\\
\noalign{\hrule height 0.5pt}
RMSE [dB]       &  0.391 & 1.337 & 0.406 & 0.547 & 0.572\\ 
Std. dev. [dB] 	&  0.173 & 0.523 & 0.084 & 0.206 & 0.066\\
\noalign{\hrule height 1pt}
\end{tabular}
\end {table}
\egroup	
	
\begin{figure}[!t]
	\centering
	\vspace{-5pt}
		\includegraphics[width=0.65\linewidth]{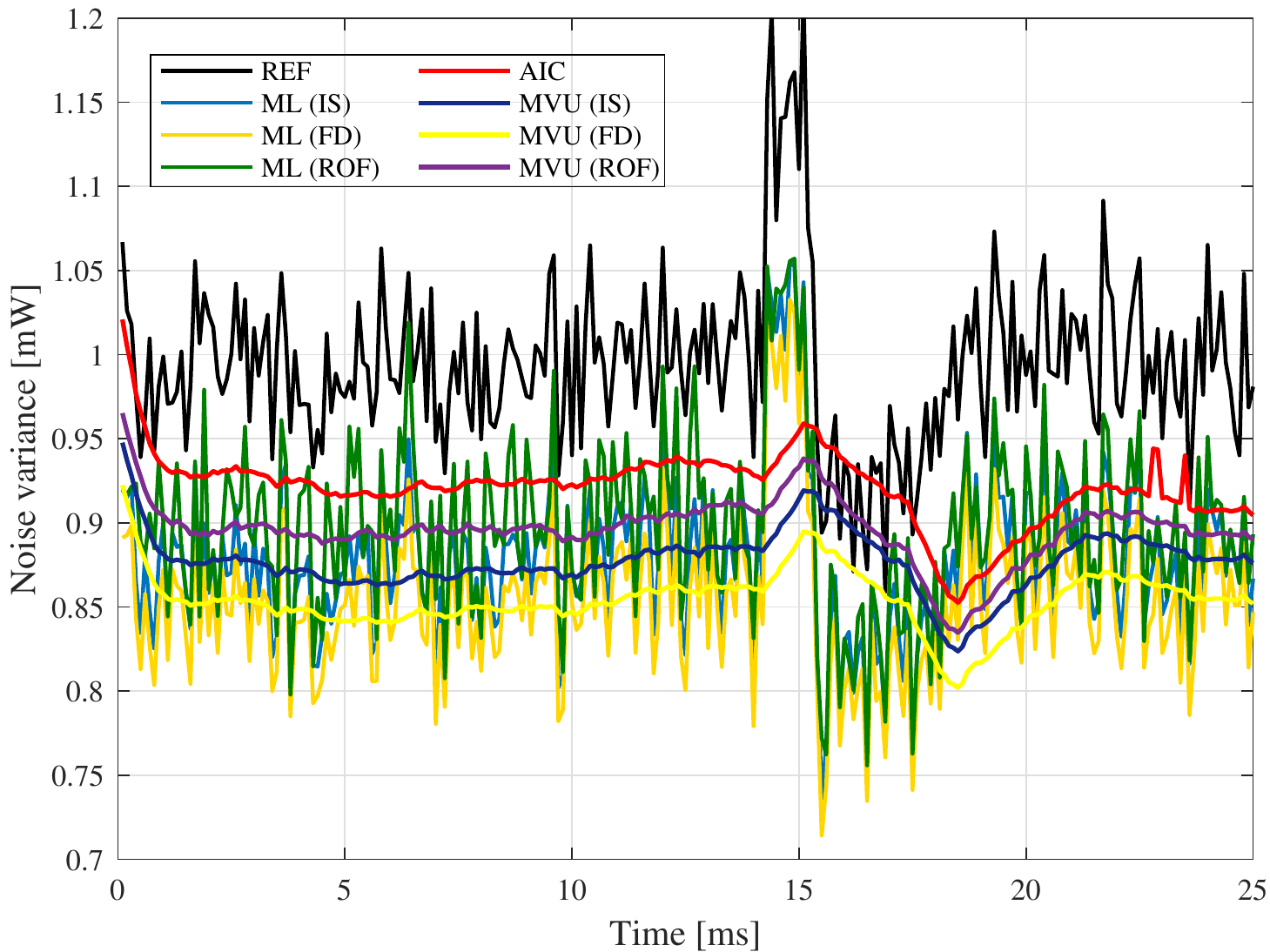}
		\vspace{-12pt}
	\caption{Comparison of the noise power estimation for different signal separation methods.}
	\label{fig:NPE_SM}
	%\vspace{-14pt}
\end{figure}
\begin{figure}[!t]
	\centering
	\vspace{-5pt}
		\includegraphics[width=0.65\linewidth]{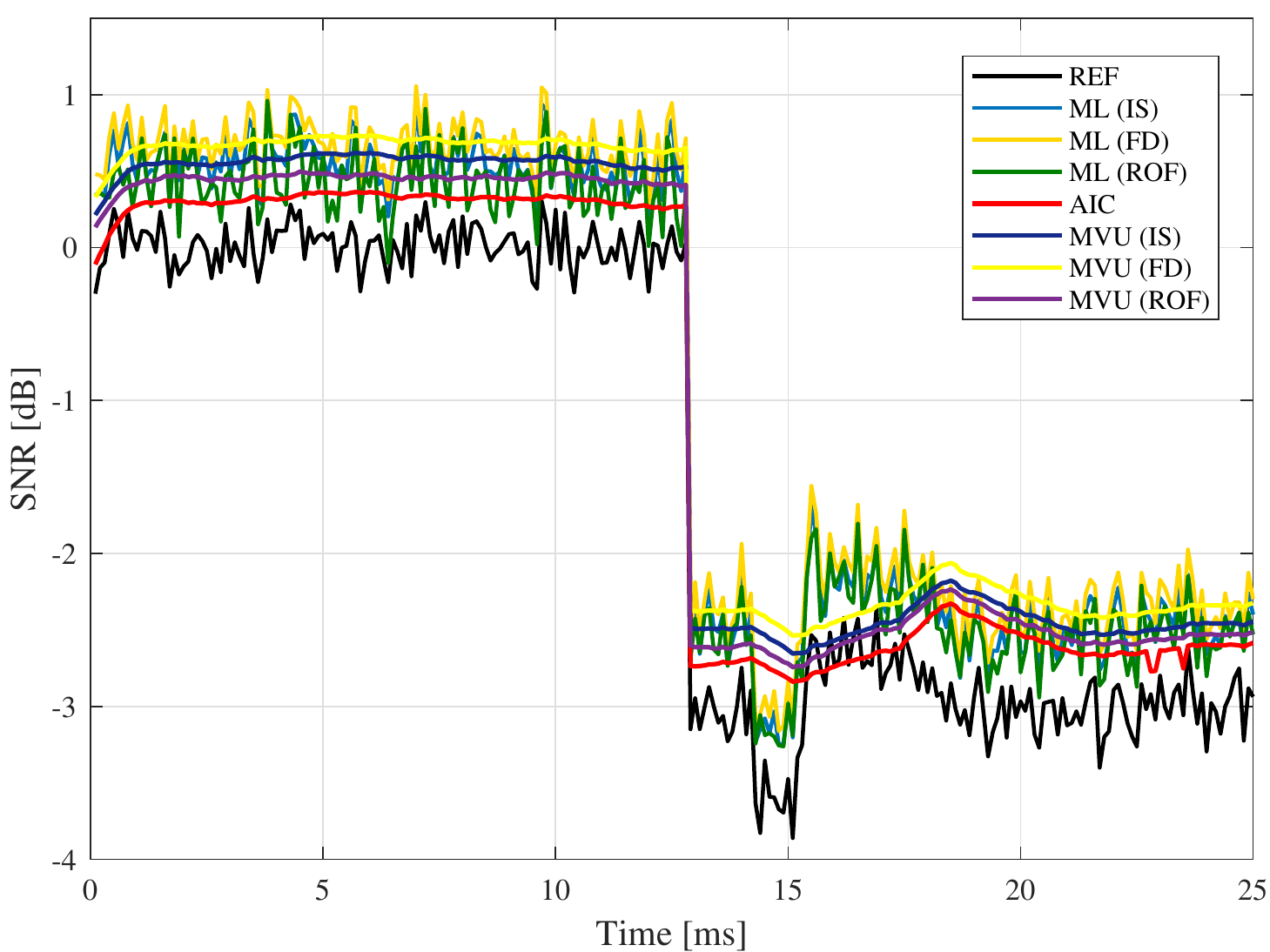}
		\vspace{-12pt}
	\caption{Comparison of SNR estimation for different signal separation methods.}
	\label{fig:SNR_SM}
	%\vspace{-14pt}
\end{figure}
\bgroup
\def\arraystretch{1.3}%  1 is the default, change whatever you need
\begin{table}[!h]
%\small
\centering
\vspace{-10pt}
 \caption{SNR Estimation Accuracy and Stability for Different Signal Separation Methods}
\vspace{-10pt}
\label{tab:Tab2}
\centering
\begin{tabular}{l|cccc}
\noalign{\hrule height 1pt}
 \textbf{Parameter} & \textbf{AIC} & \textbf{MVU(ID)} & \textbf{MVU(FD)} & \textbf{MVU(ROF)}\\
\noalign{\hrule height 0.5pt}
RMSE [dB]       &  0.406	& 0.572	& 0.706	& 0.497\\ 
Std. dev. [dB] 	&  0.084	& 0.066	& 0.064	& 0.073\\
\noalign{\hrule height 1pt}
\end{tabular}
\end {table}
\egroup

On the other hand, ML effectively tracks the current noise power. In terms of 
power regulation, the use of ML compared to CBE, will result in small but sudden 
power changes. However, due to increased inaccuracy, it exhibits higher SNR 
overestimation than eigenvalue based method. 
	
MVU significantly decreases susceptibility for short-term changes. Although 
in terms of stability it exhibits the results comparable to those achieved by 
AIC, in accuracy it causes 50\% higher SNR overestimation.

It should be noted that despite the overestimation of SNR, both ML and MVU 
remain very flexible solutions. Although in the fast processing cases it can 
be more convenient to use only ML estimator, in more complex cases method can 
be simply extend to more reliable MVU.

As we mentioned in Section~\ref{sec:Intro}, it is necessary to use signal 
separation techniques for noise power estimation using ML and MVU. 
In such a case, it is reasonable to assume that the 
problem of limited accuracy can be mitigated not only by more advanced 
estimation methods, but also by precise detection of spectrum 
parts free from deterministic signal.

We analyze ML and MVU approximations with different 
separation methods in terms of noise power estimation (Fig.~\ref{fig:NPE_SM}) 
and correctness of SNR tracking (Fig.~\ref{fig:SNR_SM}). ML and MVU methods are combined with 
ideal separation (IS), Fisher discriminant (FD) and proposed rank order 
filtering (ROF). MDL implementation is omitted, due to its similarity 
with the AIC method \cite{Sequeira2012On}. The AIC is observed to be the most competitive 
estimation method in terms of accuracy and complexity of estimation. 
Therefore it is used as the main reference technique for the 
proposed solution.

The ideal splitting case exhibits an underestimation effect because of the 
reduced size of the noise group compared to the reference and underestimation 
resulting from inaccurate transition to the frequency domain. The effect 
increases with the use of Fisher's discriminant. Due to the recognition of high 
amplitude noise samples as part of the signal group, the use of discriminant 
results in the highest SNR overestimation in the analyzed group of methods.

In the presented case, ROF results in the opposite effect i.e., a lower 
underestimation. As the signal 
samples with near-noise values at signal boundaries are not rejected, it 
results in the extension of the noise group by a small number of low signal 
samples and improves the noise power underestimation. The proposed solution 
provides better sample separation and as a consequence in conjunction with 
ML and MVU improves SNR estimation (see Table~\ref{tab:Tab2}). With the small 
cost 
of decrease in stability, the use of ROF significantly improves the accuracy 
of  
long-term response to SNR changes. The proposed solution clearly outperforms 
use of Fisher discriminant in terms of accuracy and remains a convenient 
alternative to AIC.

%%%%%%%%%%%%%%%%%%%%%%%%%%%%%%%%%%%%%%%%%%%%%%%%%%%%%%%%%%%%%%%%%%%%
%%%%%%%%%%%%%%%%%%%%%%%%%%%%%%%%%%%%%%%%%%%%%%%%%%%%%%%%%%%%%%%%%%%%
%%%%%%%%%%%%%%%%%%%%%%%%%%%%%%%%%%%%%%%%%%%%%%%%%%%%%%%%%%%%%%%%%%%%
%%%%%%%%%%%%%%%%%%%%%%%%%%%%%%%%%%%%%%%%%%%%%%%%%%%%%%%%%%%%%%%%%%%%
%%%%%%%%%%%%%%%%%%%%%%%%%%%%%%%%%%%%%%%%%%%%%%%%%%%%%%%%%%%%%%%%%%%%
\section{Computational Complexity}
\label{sec:CompComplexity}

To complete the comparison it is necessary to determine the computational 
complexity of each estimation method. Assuming the basic operations performed in IEEE 
arithmetic machine at $O(1)$ cost each, complexity is defined as the 
number of operations necessary to perform detection on $N$ received new samples. 
To simplify the calculation, an $N\times N$ block size is adopted. Each estimation 
method precedes FFT at cost $N\log N$.

To perform covariance based estimation (CBE) on $N$ incoming samples, it 
is necessary to perform matrix multiplication with $N^{2.38}$ operations. Then, 
CBE requires determination of eigenvalues with cost of $N^2$ calculations 
assuming power iteration algorithm. The goodness of fit testing with MP distribution has a complexity of $N^2$. The algorithm requires further $8 N$ additions and 
multiplications. Thus, the total cost of the estimation will be $N\log N + 8N + 2
N^2+N^{2.38}$.

MMSE estimation requires an update of the variance vector at cost $3N^2+2N$ and 
update of the autocorrelation at cost of $N\log N$. Based on the autocorrelation, a sample 
covariance matrix is written at cost $2N$. Calculation of the power spectrum 
requires another $5N$ operations, while determining the coefficients for the 
weighted summation costs $3N^2$ operations. The total computational complexity 
of the estimation is therefore $2N\log N + 8N + 6N^2$.

In case of AIC, calculation proceeds with $7N$ operations to obtain the 
average periodogram. Sorting requires further $N\log N$ operations. The $2N^2+4N$ 
calculations are necessary to determine the function $\alpha$. Calculating AIC 
costs another $6N$. Finding a minimum of the function and calculating the 
power from the remaining samples of the periodogram will cost maximum $2N$ 
operations. The total cost of AIC estimation will therefore be $2N\log N + 17N +2
N^2$.

Determining the ML or MVU estimates for a single $N$ samples vector with 
respect to the Fisher discriminant will require firstly $4N$ operations to 
determine the amplitude spectrum. Finding the maximal Fisher's discriminant 
involves the cost of $4N$ each time it needs to be designated, whereas the 
discriminant can be found in $N$ steps. This gives the cost of $4N^2$ operations. 
Determination of power on the basis of separated samples in the maximal case 
will require $2N$ operations. Thus, the total cost of the estimate is $N\log N+6N+4
N^2$.

In the proposed algorithm, that is merging ML or MVU with ROF, it is 
necessary to determine the power spectrum at the expense of $5N$ operations. 
Filtering process and power calculations require further $2N^2$ operations. 
Later, the algorithm is loaded with the sum of $5N$ operations resulting from 
calculating differences in powers, comparisons to find desired power drop 
point, moving average, derivative and samples rejection. The cost of 
conducting ML on the remaining power spectrum samples will require maximally $
N$ operations. The total cost of the proposed estimation technique will be $N
\log N + 11N + 2N^2$.

The comparison of computational complexity in Big-O notation shows for most 
methods the same $O(n^2)$ square complexity. However a deeper analysis of all 
necessary operations (see Table~\ref{tab:Tab3}), shows that the proposed method exhibits lower 
complexity with respect to competing solutions. These small computational 
differences are of particular importance in cases of frequently performed 
fast-sensing based on a small number of input samples.

\bgroup
\def\arraystretch{1.3}%  1 is the default, change whatever you need
\begin{table}[!t]
%\small
\centering
 \caption{Computational Complexity of Studied Estimation Methods For $N$-Samples Input}
\label{tab:Tab3}
\centering
\begin{tabular}{ll}
\noalign{\hrule height 1pt}
 \textbf{Estimation method} & \textbf{Computational complexity}\\
\noalign{\hrule height 0.5pt}
CBE       &  $8N +N\log N + 2N^2+N^{2.38}$\\
MMSE 	    &  $8N + 2N\log N + 6N^2$\\
AIC 	    &  $17N + 2N\log N + 2N^2$\\
ML (FD) 	&  $6N + N\log N + 4N^2$\\
ML (ROF) 	&  $11N + N\log N + 2N^2$\\
\noalign{\hrule height 1pt}
\end{tabular}
\end {table}
\egroup

%%%%%%%%%%%%%%%%%%%%%%%%%%%%%%%%%%%%%%%%%%%%%%%%%%%%%%%%%%%%%%%%%%%%
%%%%%%%%%%%%%%%%%%%%%%%%%%%%%%%%%%%%%%%%%%%%%%%%%%%%%%%%%%%%%%%%%%%%
%%%%%%%%%%%%%%%%%%%%%%%%%%%%%%%%%%%%%%%%%%%%%%%%%%%%%%%%%%%%%%%%%%%%
%%%%%%%%%%%%%%%%%%%%%%%%%%%%%%%%%%%%%%%%%%%%%%%%%%%%%%%%%%%%%%%%%%%%
%%%%%%%%%%%%%%%%%%%%%%%%%%%%%%%%%%%%%%%%%%%%%%%%%%%%%%%%%%%%%%%%%%%%
\section{Conclusion} 
\label{sec:Conclusion} 
In this paper, we performed a comprehensive analysis of noise power 
estimation methods in terms of their stability, accuracy and complexity. In 
our analysis, a special attention is given to the background noise process to 
obtain a real-world performance of the existing estimation techniques. For 
this purpose, we utilized complex noise measurement data collected in an 
industrial plant. We observed that the measured noise process has completely 
different behavior on the performance of the studied estimators when compared 
with pseudorandom noise. 

Our analysis reveals that exploring the new solutions for effective noise 
power estimation techniques is one side of the coin. The other side is to 
find the efficient methods of samples separation for prior estimation 
techniques. In this respect, we developed a novel algorithm for samples 
separation that finds its roots in rank order filtering. The proposed 
algorithm, while being simple to implement thanks to its low-complexity, is 
effective in improving the accuracy of noise power and SNR estimation when 
combined with the commonly used ML and MVU estimation algorithms. Our 
proposal, results in less than 0.5 dB RMSE and 0.075 dB standard deviation, 
respectively. The achieved performance makes our method comparable to more 
complex solutions based on covariance matrix analysis, and also the ones 
exploiting information theory concepts.

%% use section* for acknowledgment
\section*{Acknowledgment}
The authors would like to thank Simone Grimaldi for his help in collecting  
noise traces from an industrial site.

\ifCLASSOPTIONcaptionsoff
  \newpage
\fi


\bibliographystyle{IEEEtran}
\bibliography{TIMbib}

\end{document}